# A Roadmap Towards Building a National Smart Campus: An Ecosystem Approach

Larry Abdullai
*dept. of software engineering*
*LUT University*
Lappeenranta, Finland
larry.abdullai@lut.fi

Jari Porras
*dept. of software engineering*
*LUT University*
Lappeenranta, Finland
jari.porras@lut.fi

Md Sanaul Haque
*dept. of software engineering*
*LUT University*
Lappeenranta, Finland
md.haque@lut.fi

*Abstract*—Universities are racing towards making their campuses and cities smart in response to the global digitalization trend. However, the sustainability impact of Smart Campus' research, development, and innovation services on other relevant stakeholders such as the small and medium-sized businesses, remain under-investigated. The Finnish National Smart Campus project seeks to bridge this gap by orchestrating a SC ecosystem where eight SC collaborate to bring trailblazing services to businesses and society. To maximize the sustainability impact of the SC ecosystem, this study used a participatory workshop to identify the challenges of SC, provide a step-by-step guide on how to identify other relevant stakeholders, and ascertain the perceived sustainability impact using one of the SC ecosystem's RDIs as a case study. The preliminary results revealed that barriers to university-industry ecosystem development include (i), the lack of clarity in the shared goals (i.e., value proposition) between actors and (ii), weak stakeholder involvement in university RDI processes. Finally, this paper proposed a SC ecosystem model which offers a mindset shift for higher educational institutions in promoting the convergence of SC services and sustainability to support the sustainable development of Finnish-based SMEs.

*Keywords— smart campus, sustainability, ecosystem, impact, stakeholders*

## I. Introduction

The word "smart" has become synonymous with anything that can exhibit elements of intelligence. Smart campus (SC) is an emerging concept among researchers and policymakers [1]; hence, its definition comes in different shapes and forms based on, for instance, its technological component [2], [3], resemblance to smart cities [4] and based on its intelligence and functions [5], [6]. Other studies have also identified and defined the characteristics of SC [4], [7], [8]. Until now, most of the existing SC research, development and innovations (RDIs) seem to fulfil the needs of a section of stakeholders such as students, parents, University teaching and non-teaching staff [9], [10]. Scholars posit that SC can serve as a powerful platform to bridge the gap between universities, businesses and society in tackling complex sustainability challenges that face humanity today [11], [12]. The challenge however is redefining the value proposition (VP) of SC to maximize the sustainability impact of its RDIs on broader stakeholders. Based on existing features of SC as shown in Fig.1, we argue that applying an ecosystem thinking could extend the sustainability impact of SC to other relevant stakeholders. For instance, how can the sustainability impact of elements such as virtual reality (VR) application in Fig. 1 be extended to other stakeholders? The aim of the study is therefore to explore the challenges and sustainability impact of SC RDIs on especially businesses and society.

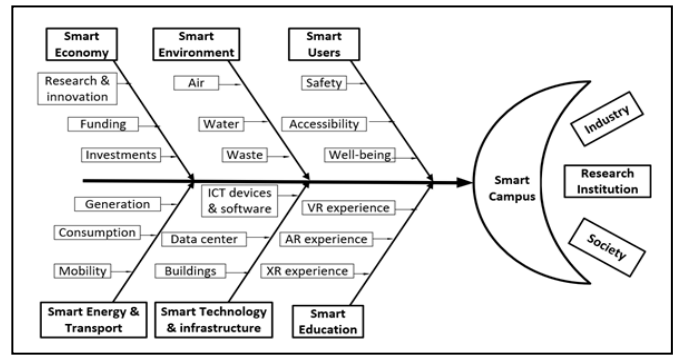

Fig. 1. Fishbone model of smart campus domain. (*Source: Authors' own*)

## II. Methodology

An exploratory single case study [13] has been applied to conduct the study. Although single case study provides a narrow scope of generalizability, yet it provided an in-dept focus which is replicable. Data was collected through a participatory workshop that was performed via virtual platforms on the 1st of November 2021 with Tampere Open Public Transport Service (TOPTS) as the case study. There were 8 participants representing different partners in the Smart Campus project. The workshop utilized semi-structured questions to gain insight into TOPTS' VP and challenges, identify relevant stakeholders, and conduct a sustainability impact assessment of TOPTS to the identified stakeholders using the Karlskrona sustainability awareness framework [14]. The workshop consisted of three parts focusing on:

*1) Challenges and benefits: Using a value proposition canvas to ascertain TOPTS' benefits and challenges*

*2) Stakeholder brainstorming: To identify, categorize, map and select relevant stakeholders for TOPTS,*

*3) Sustainability impact: To assess the perceived sustainability impact of TOPTS on the other stakeholders.*

### B. The Finnish National Smart Campus Ecosystem Project

The Finnish National Smart Campus Ecosystem FNSCE project was started in early 2021 by eight Finnish Universities and the University of Applied Sciences. The strength of the FNSCE is its capacity to accelerate the upscaling of trailblazing sustainable research-based innovations and make them effortlessly accessible to Finnish-based SMEs while shortening the research to industry journey. The SC RDIs of which TOPTS is part, rely heavily on advanced edge computing, sensors, AR/VR and 5G/6G wireless technologies and are expected to create significant societal impact by responding to complex societal and business challenges with novel applications and services.

## C. Case Study: TOPTS

Tampere Open Public Transportation Service, as shown in Fig. 2, is one of the SC RDI services led by Tampere University. TOPTS is built on a software called Virrake, a VR immersive environment that aims at addressing transport accessibility issues for vision, hearing or physically impaired users. This case was selected to explore the challenges in conveying the value of SC RDI services to businesses and society and to create awareness of the potential sustainability impact of SC RDI services.

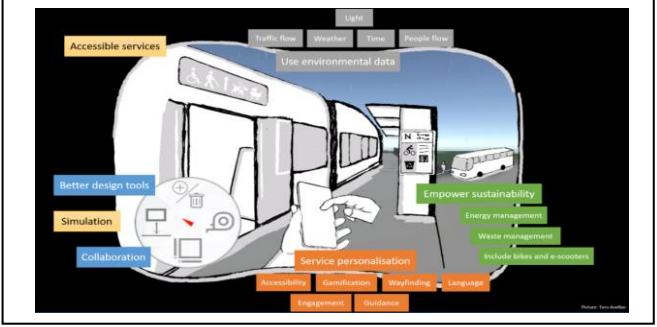

Fig. 2. Application of Virrake to the open public transport service [15].

## III. RESULTS

The key findings from the study were the perceived positive sustainability impact of TOPTS as summarized in Table 1. However, the most noteworthy insight from the workshop was that the main barriers to university-industry-society ecosystem development include: (i) the lack of clarity of SC value proposition to industry and society, and (ii) weak stakeholder involvement in SC RDI processes. Participants also mentioned challenges such as "misunderstanding", "trust", and "RDI not meeting the needs of stakeholders" as reasons why stakeholders might not be willing to engage.

TABLE I. SUSTAINABILITY IMPACT OF THE CASE

| Sustainability Dimensions | Impact |
|---|---|
| Economic | (1) Development of local business (2) More profit (3) No extra cost (4) Increased use of public transport (5) It is scalable |
| Social | (1) People spend more time outside and together (2) Ease of social communication (3) Easy to report problems (user friendly) (4) Special engagement with the city transport planning (5) Person is seen as "digi native" |
| Environmental | (1) Increased use of public transport (2) Smart and sustainable city (3) Moderate energy consumption (4) Clean and fresh air in the city (no air pollution) (5) Leads to decreased use of private cars |
| Technical | (1) Secure (2) It is scalable (3) Fast (4) Easy to adapt to an existing service (5) Easy to maintain |
| Individual | (1) Ability to choose between different transportation (2) Feeling secure (3) Trust (4) Comfortable transportation (5) Personalized transportation support (6) Faster transportation (7) User-friendly transportation |

## IV. SMART CAMPUS ECOSYSTEM MODEL

In this section, we discuss another approach to viewing SC through the ecosystem lens based on the workshop results. Adner defines an ecosystem as "the alignment structure of the multilateral set of partners that need to interact for a focal value proposition to materialize" [16]. Thus, the ability of SC to co-create value with other stakeholders and be part of a strong ecosystem becomes a potential and powerful source of competitive advantage. To achieve this, we merged the concepts of smart campus, sustainability, and ecosystem to propose a smart campus ecosystem model which is described in the subsequent paragraph.

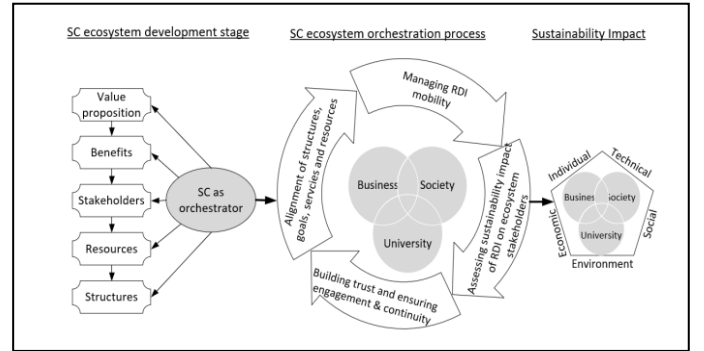

Fig. 3. Smart campus ecosystem model.

## A. Smart campus ecosystem development stage

This component of Fig.3 represents the ecosystem elements based on [16] ecosystem definition. VP represents the SC RDI services that promises benefits to stakeholders. These stakeholders contribute resources which must be aligned with the elements to maximize the sustainability impact of the VP. As such, a clear VP forms the foundation for stakeholder engagement and sustainability [17]

## B. Smart campus orchestration process

The second component of the model shows the deliberate and purposeful activities necessary to materialize the VP. It includes activities such as coordination, mobilization of resources and stimulating sustainable innovation.

## C. Sustainability impact

Finally, the model includes a sustainability impact assessment phase to maximize the sustainability impact of the SC ecosystem services on the stakeholders which comprise of three major players: the university, industry, and society. As shown in Table I, extending SC RDI service to other stakeholders led to a better city planning and increased use of public transport especially among the underserved community. This also resulted in better air quality in the city.

## V. DISCUSSION AND CONCLUSION

To conclude, the workshop aided the participants to clarify the VP of TOPTS, identified other relevant stakeholders to be engaged with and brainstormed on potential sustainability impact of TOPTS on stakeholders. The results indicated that clarity in conveying the SC VP to industry and society is essential for stakeholder engagement and value co-creation [18]. Since individual stakeholders differ largely in their interest to the VP [19], we argue that the proposed SC ecosystem model (Fig.3) could provide a pathway for SC to clarify their VP and engage with relevant stakeholders for sustainability [17] and value co-creation. This study contributes to the understanding of the sustainability impact of SC initiatives and ecosystem thinking.

## ACKNOWLEDGMENT

This study is supported by the Finnish Academy funded project Smart Campus National RDI Network – From Innovation to Business. We are indebted to Tampere University team for their immense contribution to the workshop.